\documentclass[traditabstract]{aa} 
\usepackage{graphicx}
\usepackage{txfonts}
\usepackage{wasysym}
\usepackage{natbib}
\bibpunct{(}{)}{;}{a}{}{,} 
\usepackage{ulem}
\begin{document}
   \title{An HST search for planets in the lower main \\ sequence of the globular cluster NGC 
6397\thanks{Based on observations with the NASA/ESA Hubble Space Telescope, obtained at the Space Telescope
Science Institute, which is operated by AURA, Inc., under NASA contract NAS 5-26555.}}

   \author{V. Nascimbeni\inst{1,2,3}\thanks{Visiting PhD Student at STScI (DDRF D0001.82432
           program).}
          \and L.\ R. Bedin\inst{2,3}
          \and G. Piotto\inst{1,2}
          \and F. De Marchi\inst{4}
          \and R. M. Rich\inst{5}
          }

   \institute{Dipartimento di Astronomia, Universit\`a degli Studi di Padova,
              Vicolo dell'Osservatorio 3, 35122 Padova, Italy\\
              \email{valerio.nascimbeni@unipd.it, giampaolo.piotto@unipd.it}
         \and
             INAF -- Osservatorio Astronomico di Padova, vicolo dell'Osservatorio 5, 35122 Padova, Italy\\
             \email{luigi.bedin@oapd.inaf.it}
         \and
             Space Telescope Science Institute,
             3700 San Martin Drive, Baltimore, MD 21218
         \and Dipartimento di Fisica, Universit\`a di Roma Tor Vergata and INFN, Sezione
              di Roma Tor Vergata, I-00133 Roma \\
             \email{fabrizio.demarchi@roma2.infn.it}
  \and
  Division of Astronomy and Astrophysics, University of California, Los Angeles, 430 Portola Plaza, Box
 951547, \\ Los Angeles, CA 90095-1547, USA.  \email{rmr@astro.ucla.edu} 
             }

   \date{Submitted December 16, 2011; accepted February 16, 2012 }

\abstract{Searches for planetary transits carried out in open and
  globular clusters have to date yielded only a handful of weak,
  unconfirmed candidates. These results have been interpreted as either 
  being insignificant, or evidence that the cluster chemical or
  dynamical environment inhibits planetary formation or survival.
  Most campaigns have been limited by small sample statistics or
  systematics from ground-based photometry.  We
  performed a search for transiting planets and variables in a deep
  stellar field of NGC 6397 imaged by HST-ACS over 126 orbits.  We
  analyzed 5,078 light curves, including a careful selection of 2,215
  cluster-member M0--M9 dwarfs.  The light curves were corrected
  for systematic trends and inspected using several tools. No
  high-significance planetary candidate is detected.  We compared
  this null detection with the most recent results from Kepler,
  showing that no conclusive evidence of lower planet incidence can be
  drawn.  However, a very small photometric jitter 
  is measured for early-M cluster members ($\lesssim 2$ mmag on 98\% of them), which may be worth 
  targeting in the near future with more optimized campaigns.  Twelve
  variable stars are reported for the first time.  }

   \keywords{techniques: photometric -- stars: planetary systems -- clusters: individual: NGC 6397}
   \authorrunning{Nascimbeni et al.}
   \titlerunning{An HST search for planets in the lower Main Sequence of the globular cluster NGC 6397}
   \maketitle

\section{Introduction}

More than seven-hundred extrasolar planets are known
(\texttt{exo\-planet.eu} database).  Most of them are characterized
only through radial velocities (RV) and lack any information about
their ``real'' mass $M_\mathrm{p}$, because  only $M_\mathrm{p}\sin i$
can be measured, where $i$ is the inclination of the orbit with
respect to the line of sight.  Their size is also unknown, which
prevents us from getting any clues about their density and physical
composition.  Exoplanetary transits are highly complementary to RV
techniques, providing the planetary radius $R_\mathrm{p}$ from the
stellar  radius $R_\star$ in a direct geometrical way
(\citealt{seager2011}, p. 55). Photometric searches for transits can
also go much deeper than RVs in magnitude, and can monitor thousands
of stars simultaneously. The Kepler mission \citep{borucki2010}, for
instance, has demonstrated the power of this technique  by discovering
many planetary systems with unexpected properties.  Hundreds of Kepler
``candidate planets'', for which confirmation and mass measurement via
RVs remains infeasible, are still very useful for statistical purposes
\citep{howard2011,schlaufman2011}

Star clusters, and in particular globular clusters (GC), offer a
unique opportunity to study how the chemical and dynamical
environments affect planetary formation and evolution.  They are also
comprised of stars that share (in most cases)  the same age and
chemistry,  and whose radii $R_\star$ and masses $M_\star$ are
reliably known on their main sequence (MS).   Open clusters (OC) have
been targeted for extensive transit searches \citep[most
  notably]{mochejska2005,montalto2007,montalto2011,hartman2009} but
only a handful of weak, unconfirmed candidates have been so far
detected  \citep{mochejska2006,montalto2011}.  A global reanalysis
suggests that the overall statistical significance of these campaigns
is so low that it could be compatible with the planet host incidence
observed in the field \citep{vansaders2011}.  On the other hand, GCs
are on average much richer in stars than OCs, providing a much higher
statistical significance in the case of a null detection.  The only
GCs monitored for transits have been 47 Tucanae for both 8 days with
HST WFPC2 \citep{gilliland2000} and 33 days from the ground
\citep{weldrake2005}; and $\omega$ Centauri with a 25-day ground-based
campaign \citep{weldrake2008}.  No planetary detection has been
claimed, and \citet{gilliland2000} concluded that the planetary
occurrence in 47 Tuc is smaller by a factor of ten than in field
stars.  The reasons that have mostly been hypothesized to explain the
lack of giant planets in GCs are their metallicity and their dynamical
environment.

It has long been known that metallicity is a strong primary parameter
that correlates with the fraction of stars with planets
$\Phi_\mathrm{p}$.  For giant planets, \citet{fischervalenti2005},
among others, measured an increase from the typical value
$\Phi_\mathrm{p}^\odot\sim 0.03$ for stars with solar metallicity up
to $\Phi_\mathrm{p}\sim 0.25$ for very metal-rich
($[\textrm{Fe}/\textrm{H}]\gtrsim +0.3$)  stars.   For moderately
metal-poor stars ($-0.5\lesssim [\textrm{Fe}/\textrm{H}]\lesssim 0$),
it is still disputed whether this  correlation becomes flat at values
$\Phi_\mathrm{p}\simeq \Phi_\mathrm{p}^\odot$
\citep{udry2007,santos2011} or $\Phi_\mathrm{p}$ continues to decrease
exponentially towards lower metallicities \citep{johnson2010}.  As
more and more low-mass planets ($M_\mathrm{p}\lesssim 30 M_\oplus$)
are being discovered, it becomes clear that their $\Phi_\mathrm{p}$ is
much larger than that of giant planets
\citep{lovis2009,wittenmyer2011}, probably around $\sim$20--30 \%.
The occurrence of low-mass planets seems to be insensitive to the host
star  metallicity \citep{udry2006}, except maybe for very late-type
stars \citep{johnson2009,schlaufman2011} but the M-dwarf metallicity
calibration is still too uncertain to draw conclusions.

The complex dynamical environment of a cluster is the second major
concern  about the survival of protoplanetary systems. Theoretical studies
give different answers, but  some of them imply that gravitational
stripping processes are not strong enough to disrupt very short-period
($P<5$ days) planetary systems, even in the densest regions of a
typical GC \citep{fregeau2006,spurzem2009}. Planets do indeed exist
in clusters. Giant planets have been detected around evolved stars
belonging to open clusters \citep{lovismayor2007,sato2007}, and around
a pulsar in the globular cluster M4 \citep{thorsett1999}. It is therefore 
difficult to understand why no planets have been detected in 47 Tuc.
The search by \citet{gilliland2000} was sensitive only to giant
planets ($R_\mathrm{p}\gtrsim 1 R_\mathrm{jup}$) around stars in the
upper MS, excluding late-K and M dwarfs.  More data are needed to
sample other regions of the  parameter space.

On the other hand, it is convenient to search for transits around KM
main-sequence stars because of  their larger expected signal. These
targets, even in the nearest GCs, are faint ($V>18$) and extremely
crowded. Space-based observations are necessary to achieve the needed
signal-to-noise ratio (S/N)  per time unit, and to minimize the number
of false positives caused by blended  photometric contaminants.  The
wide-field imagers mounted on HST (ACS and WFC3) are unrivalled  in
this respect, and their large archive of  deep photometric series of
GCs is already available to be exploited for  transit
searches. However, we show in Sec. \ref{sec:zp} that  HST
time-resolved photometry is influenced by many sources of systematic
errors that require a careful correction. We developed specific tools
to this purpose and applied them to a test case.  Other data sets
(such as those for 47 Tuc) will be investigated in the near future, to
test whether a more optimized campaign is worth  pursuing.

In this paper, we present a search for variables and  planetary
transits in NGC 6397, exploiting a 126-orbit data set from the HST
Advanced Camera for Surveys (ACS; GO-10424; PI:Richer).   NGC 6397 is
a metal-poor ($[\textrm{Fe}/\textrm{H}]\simeq -2$) core-collapsed
globular cluster, and the second-closest known
\citep{gratton2003,richer2008}.   Our work is in some sense
complementary to that presented by \citet{stello2009}, who employed
the same data set to study the microvariability of highly saturated
red giants.  Though these data were not optimized for a transit
search,  its unprecedented depth allowed us to search for planets on a
homogeneous sample of 2,215 member  M-dwarf stars down to the
hydrogen-burning limit. The M dwarfs are considered the most promising
targets to discover rocky planets in the near future
\citep{scalo2007}.  They are the smallest stars ($R_\star \simeq
0.1$--$0.5 R_\odot$), and therefore the transit depth for a given
planet is larger by a factor of $\sim4$--$100$ that that of a
solar-type host star. They are also intrinsically faint ($L_\star
\simeq 0.02$--$5\cdot10^{-5}L_\odot$), meaning that a habitable planet
would have an orbital period of only $P=10$--30 days.

\section{Observations and data reduction}

\begin{figure*}[!t]
\centering
\includegraphics[height=9.1cm]{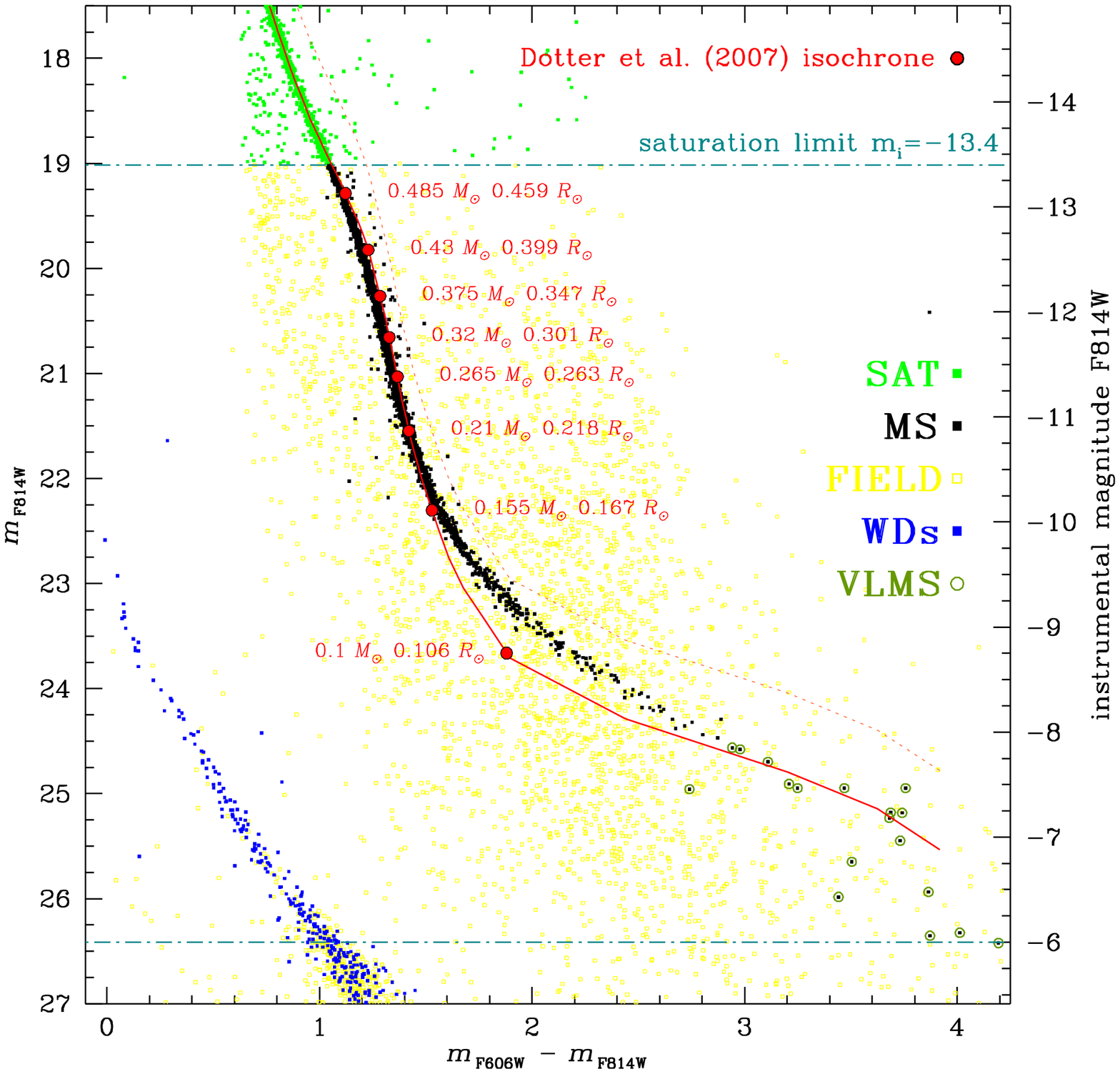}
\includegraphics[height=9.1cm]{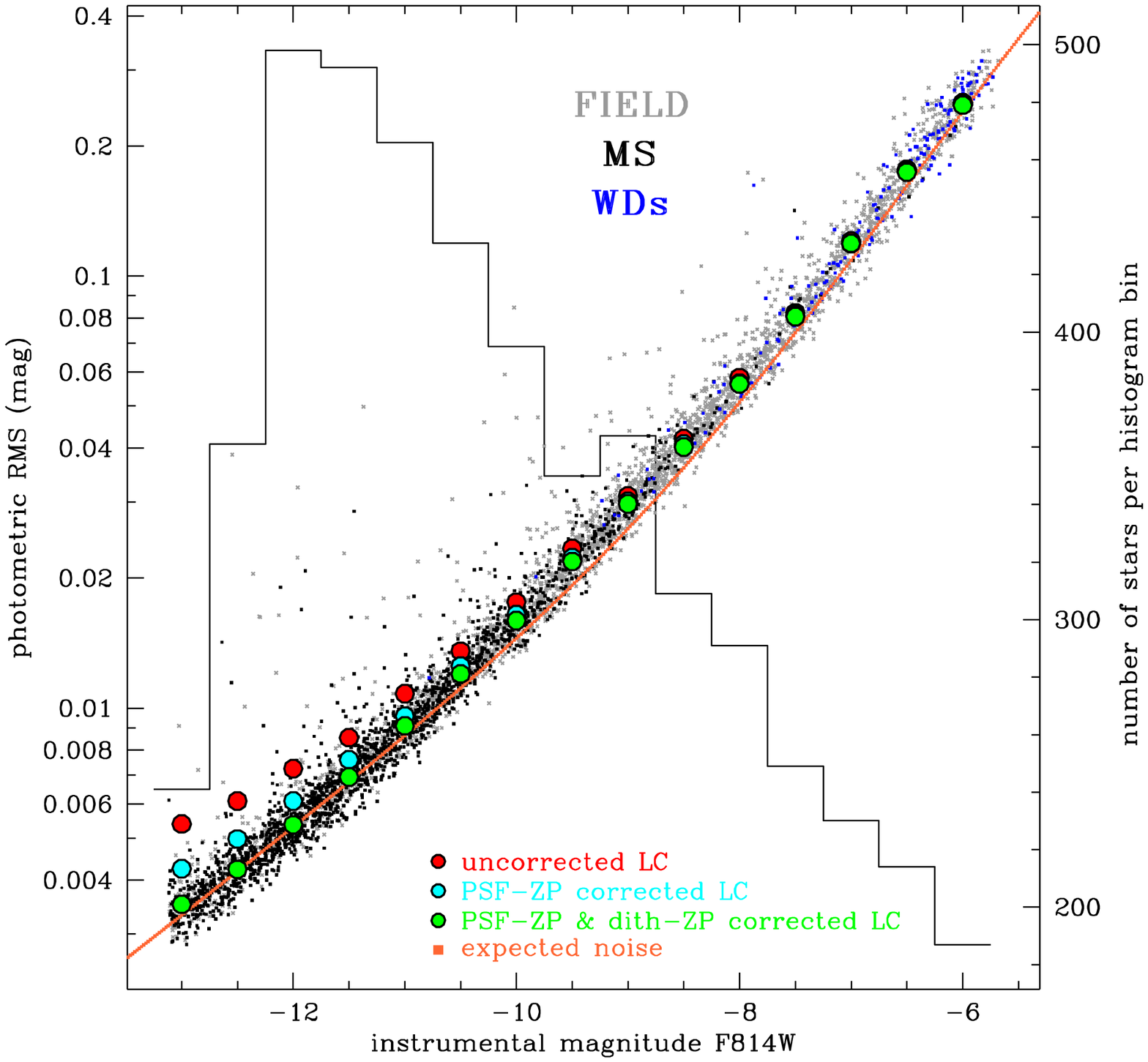}
\caption{ 
\emph{Left panel.} Color-magnitude diagram
  ($m_\mathrm{F606W}-m_\mathrm{F814W},m _\mathrm{F814W}$) for all the
  stars in the \citet{anderson2008} master list; the  5,078 sources
  selected in this work lie  between the saturation limit at
  $m_\mathrm{F814W}\simeq 19$ and the faint limit
  $m_\mathrm{F814W}\simeq 26.5$ (dash-dot lines). The red line is the
  isochrone by \citet{dotter2007} employed in \citet{richer2008}. 
  The dotted line marks the loci occupied by equal-mass MS-MS binaries.
  \emph{Right panel.} The RMS for light curves in our sample  as a function
  of the instrumental magnitude. Red
  circles: median RMS averaged over 0.5 mag bins, without any correction. 
  Cyan circles: the same, after PSF-ZP correction. Green circles: 
  both PSF-ZP and dither-ZP corrections
  applied. Small points are the RMS of the individual light curves
  after both corrections. The solid orange line is the
  expected theoretical noise level. The superimposed histogram represents
  the number of targets in each 0.5 mag bin (scale at the right). 
 }
\label{cmd}
\end{figure*}

Our analysis is based on the HST data set GO-10424 (PI: Richer), whose
acquisition was originally made to probe the bottom of the MS and the
end of the white dwarf (WD) cooling sequence in NGC 6397
\citep{hansen2007,richer2008}.  A single $202''\times202''$ field,
located $5'$ from the cluster core  was imaged for 126 orbits with the
wide-field channel of the ACS. In each orbit, a single exposure
through the F606W filter was preceded and succeeded by two F814W
images, with exposure times ranging between 584 s and 804 s (median:
704 s).  Overall, 252 F814W frames and 126 F606W frames were secured,
for a total exposure time of $\sim 50$ hours.  Shorter exposures
($t_\mathrm{exp}=1$, 5, 40 s) were also taken to measure the fluxes of
the brightest stars, but they  were not used in our study. 

The dynamic range of the ``deep'' exposures is perfect for our purposes,
as the observed luminosity function (LF) of the cluster MS members
peaks at $m_\mathrm{F814W}\simeq 21$ (that is, on well-measured stars with
S/N $\sim 200$ in the same filter; 
Fig. \ref{cmd}, histogram on the right panel). Saturation occurs around
M0V spectral type at $m_\mathrm{F814W}\simeq 18.7$ and the faintest cluster
members (M9V) are found at $m_{F814W}\simeq 24$ (S/N $\sim 20$), such that a
sample of $\sim 2,000$ cluster  M dwarfs down to the hydrogen-burning
limit is available to transit search (Fig. \ref{cmd}, left panel).

On the other hand, these data were not optimized to search for transit
events, thus two aspects of the observing setup are somewhat
restrictive. First, the time coverage of the frames is discontinuous,
unlike in \citet{gilliland2000}. The data for $\sim 50$ h of
integration time were spread among twenty-one non-contiguous days,
spanning a twenty-eight day period. This translated into a much lower
completeness for our search, especially for the longest periods ($P>3$
d, see Sec. \ref{sec:compl}). The second reason is that each pointing
was shifted within a ten-position dithering pattern $\Delta x, \Delta
y$  plus a subpixel offset $\delta x, \delta y$. While dithering is
usually suitable  for undersampled images \citep{andersonking2000},
it prevented us from reaching the highest photometric accuracy
possible for the brightest stars. For these targets, the amount of
random noise is so low that the  unavoidable flat-field and
pixel-to-pixel residual errors are no longer negligible. A zero-point
correction, discussed in the next Section, was developed to suppress
these systematic errors.

We carried out the data reduction on individual  bias-corrected and
flat-fielded \texttt{.flt} images provided by the HST pipeline.  We
employed the master input list from \citet{anderson2008}, and the code
described by  \citet{andersonking2006} based upon the effective PSF
(ePSF) approach first developed by \citet{andersonking2000}. Four
tests were performed on a subset of twenty F814W frames taken at
the same integer-pixel dithering position, to choose the best reduction
strategy among: 
\begin{enumerate}
\item Allowing spatially-constant perturbed ePSFs as described by
\citet{andersonking2006};
\item Correcting the raw frames for charge
transfer (in)efficiency (CTE) with the pixel-based algorithm proposed
by \citet{andersonbedin2010}; 
\item Using both the 1) and 2) corrections; 
\item Using neither. 
\end{enumerate}
For all the measured sources, the root mean square (RMS) values of their light curve as a
function of magnitude were compared. Methods 1-3 above provided no detectable
improvement over the fourth choice, therefore we decided not to apply
any CTE or PSF correction at this stage. Actual PSFs are both
spatially and temporally variable, and require a frame-to-frame
\emph{a posteriori} correction that is  explained in Section
\ref{sec:zp}.

Hereafter, we focus our analysis on only F814W frames for many reasons.
They are much deeper than F606W images for faint, red, low MS stars,
and less affected by PSF short-term variations as shown for this
very same dataset by \citet{andersonking2006}. The sky background is also
much lower in F814W exposures. The way in which the F814W frames were sampled
(at the beginning and the end of each 96-minute orbit) ensures that the F606W
frames are nearly useless for increasing the transit detection efficiency. In
this way, we also avoided the need to  correct for any tricky registration
of the light curves  between the two filters.

Sources beyond the saturation limit of the longest exposures
($m_\mathrm{F814W}\lesssim 19$) and sources detected in fewer than 200
frames  ($m_\mathrm{F814W}\gtrsim 26.5$) were excluded from this
study, leaving 5,078 objects including cluster members, field
stars, and a limited number of non-stellar objects. We evaluated  the
membership of each entry by performing a selection on the proper
motions between our epoch and the archival ACS GO-11633 data set (PI:
Rich),  centered on the same field. We flagged 2,430 sources as
cluster members, of which 215 belong to the white dwarf (WD) sequence
(see the CMD on Fig. \ref{cmd}, left panel). The remaining 2,215 MS
stars are red dwarfs ranging from M0V spectral type
($R_\mathrm{p}\simeq 0.5 R_\odot$, $M_\mathrm{p}\simeq 0.5 M_\odot$)
down to the hydrogen-burning limit, as confirmed by superimposing 
an isochrone by \citet{dotter2007} on the CMD, as done in
\citet{richer2008}.

\section{Systematic correction}
\label{sec:zp}

\begin{figure*}
\centering
\includegraphics[height=16cm,angle=-90]{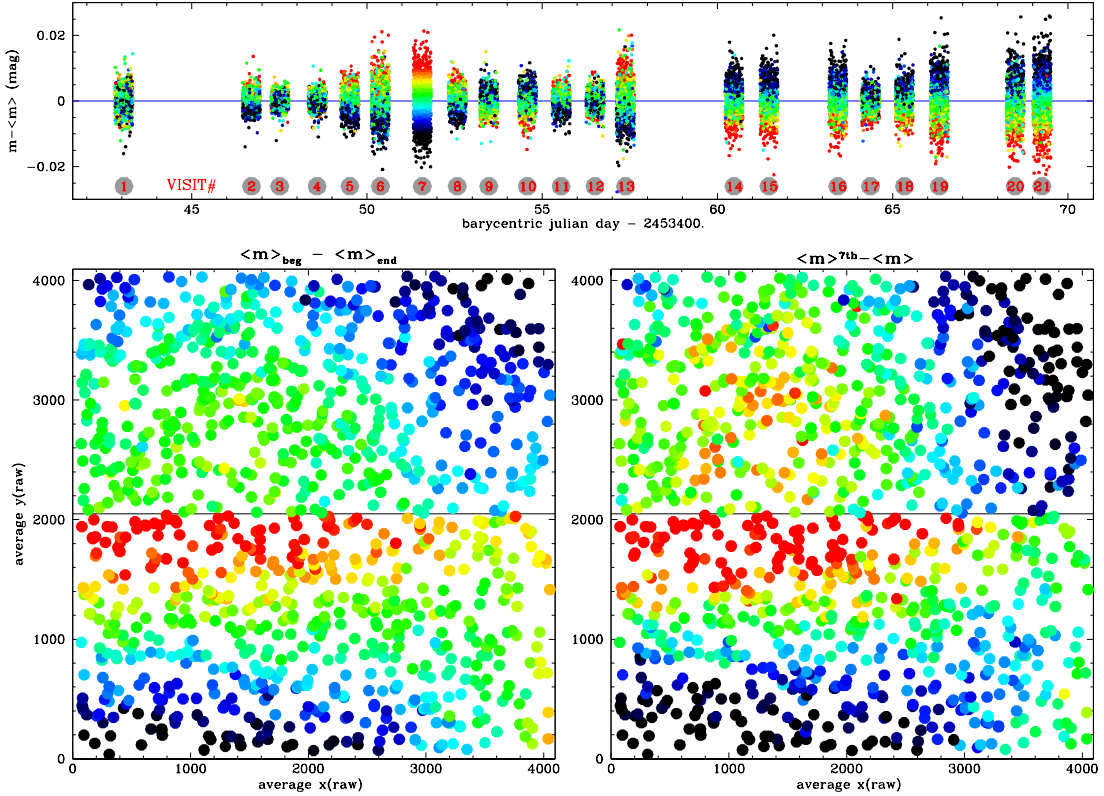}
\caption{ Mapping the PSF-ZP shift as a function of time and position on
  the ACS detector, with two different diagnostics (see text). 
  The First (\emph{left panel}) is the  difference between the median magnitudes $\langle m
  \rangle_\mathrm{beg}$ and  $\langle m\rangle_\mathrm{end}$ 
 measured in frames taken respectively at the beginning and the end of the orbit,
  while the second (\emph{right
    panel}) is the difference between the median magnitude of the star $\langle m
  \rangle^\mathrm{7th}$  measured during the seventh ``visit'' of the
  program  ($2453451<\textrm{JD}<245352$), and the median magnitude
  $\langle m \rangle$ of its full light curve. 
 \emph{Top panel:} All the high-S/N light curves ($\sigma_m<0.02$) have been 
 registered to their average magnitude $\langle m\rangle$.
 In all panels, the color scale  spans the range $-0.02$-$0.02$
  mag from black to red. 
}
\label{psfzp}
\end{figure*}

The instrumental magnitude $-2.5\log (\textrm{DN})$ of each star was
registered to the median instrumental magnitude of stars measured in
the  deepest frame of the F814W series.  We refer to this
magnitude as $m$. Saturation occurs at $m\lesssim -13.4$. 
The RMS $\sigma_m$ of our full sample of 5,078 light curves was
compared with the expected noise budget, as calculated by combining
theoretical photon-, dark-, background- and readout noises (right panel of Fig. 
\ref{cmd}, orange line). Individual $\sigma_m$ were averaged whitin
0.5 mag bins by applying a clipped median (red circles in the same plot) 
to exclude outliers from the comparison.

On the bright side, the observed noise level 
on average is far higher
than expected, even by 50-60\% for stars with $m\lesssim -12$
(red circles, right panel of Fig. \ref{cmd}). We
identified the source of most of this excess noise as a variation in the
photometric zero point (ZP) induced by systematic changes in the PSF
shape. Long-term instability  of the PSF was reported for ACS by
\citet{andersonking2006}. We noticed that this ZP change (hereafter
called \emph{PSF-ZP}) follows a well-defined pattern as a function of
time and average $x,y$ position on the detector.
The pattern can be mapped by evaluating for each
star two diagnostic parameters that appear to be strongly correlated
with the PSF-ZP shift: 
\begin{enumerate}
\item The difference between the median magnitude $\langle m
  \rangle_\mathrm{beg}$  measured in frames 
  taken at the beginning of the orbit, and the median magnitude $\langle
  m\rangle_\mathrm{end}$ measured in frames taken at the end of the orbit
  (left panel of Fig. \ref{psfzp}, color-coded in the range $-0.02$-$0.02$
  mag from black to red). 
\item The difference between the median magnitude of the star $\langle m
  \rangle^\mathrm{7th}$  measured in 16 consecutive frames taken
  during the seventh ``visit'' of the
  program  ($2453451<\textrm{JD}<245352$), and the median magnitude
  $\langle m \rangle$ of its full light curve (right panel
  of Fig. \ref{psfzp}, same scale).  
\end{enumerate}
The pattern is very similar in both cases. 
The first diagnostic $\langle m\rangle_\mathrm{beg}-\langle m\rangle_\mathrm{end}$ 
is probably a proxy of
the real  origin of the PSF-ZP systematic: a thermal/mechanical
instability linked to the orbital phase. The dependence of PSF-ZP
on time becomes evident when all the light curves with a high S/N 
(i.e., measured on all the 252 frames and having $\sigma_m<0.02$)
are registered to their average magnitude $\langle m\rangle$, stacked 
in the same plot, and color-coded as a function of 
$\langle m\rangle^\mathrm{7th}-\langle m\rangle$
(Fig. \ref{psfzp}, top panel, color scale from black to red in the
range $-0.02$ to 0.02 mag). It is clear that on average stars whose flux
is overestimated during the seventh visit are also systematically 
underestimated in the last visits (JD$>$2453460), and viceversa.

For a given frame and within the same chip, the PSF-ZP is a smooth
function of the position on the detector. The diagnostics 
$\langle m\rangle^\mathrm{7th} -  \langle m\rangle$ and
$\langle m\rangle_\mathrm{beg}-\langle m\rangle_\mathrm{end}$ 
are too noisy when evaluated for faint stars to implement an
effective correction with them.  
We chose instead to correct the
PSF-ZP with a local approach, adapted from the differential photometry
algorithms described in \citet{nascimbeni2010}. For each target star
$i$ in our sample, a set of $N$ nearby reference stars $k=1,\ldots,N$
was selected with the following criteria: 1) they had to lie on the
same chip as the target and be within 200 pixels of it; 2) they had to be 
at least 20 stars whose
total flux had to exceed ten times the flux of the target; and 3) they
had to be detected in at least 250 frames among 252, instead of the 200-frame 
limit required for a target star. When 
requirement 2) was not met, the search radius was increased until it was met. If on a
given frame $j$ a reference star $k$ was not detected,  or when its
magnitude $m_{j,k}$ was more than $3 \sigma_{j,k}$ off its value
averaged over the series  $\langle m_{j,k} \rangle$, its magnitude was
set to $\langle m_{j,k} \rangle$.  A reference magnitude $m_{0,j}$ was
calculated in each frame by performing the weighted mean  of the
magnitudes of the $N$ reference stars  $m_{0,j}=\sum_k (m_{j,k} /
\sigma^2_{j,k}) / \sum_k(1/\sigma^2_{j,k})$   \citep{broeg2005}. The
target magnitude was then normalized to $m_{0,j}$.  The PSF-ZP
correction was applied only when it decreased the overall RMS of the
target light curve. After the correction, the median RMS of the light
curves (right panel of Fig. \ref{cmd}, cyan circles) was substantially
smaller. For the brightest stars ($m\lesssim -12$), it had decreased
to a level $\sim 15$\% above the expected noise.

Most of the remaining excess noise is due to local ZP changes as well as
the dithering pattern employed: an integer $\Delta x, \Delta y$
shift plus a small sub-pixel offset  $\delta x, \delta y$ were added to the 
initial pointing $x_0, y_0$ (where the units are in physical pixels).  We refer to
this systematic effect as \emph{dith-ZP}. To each light curve, already
corrected for PSF-ZP, we applied two decorrelating algorithms where 
\begin{enumerate}
\item The median magnitude of the star $\langle m
  \rangle_{\Delta x, \Delta y}$ was calculated for each subset of frames sharing the
  same integer-pixel dither $\Delta x, \Delta y$. The magnitude $m$ of
  each frame in the subset $\Delta x,\Delta y$ was then registered to
  $\langle m \rangle_{\Delta x,\Delta y}$.
\item For each $\Delta x,\Delta y$ subset corrected by the previous
  step, we considered the magnitude $m_j$ in each frame $j$  as  a function of
  the sub-pixel shift $\delta x_j,\delta y_j$.  An ordinary
  linear least squares fitting was carried out to find the coefficients 
  $c_0$, $c_x$, and $c_y$ such that
\begin{equation}
m_j = c_0 + c_x \,\delta x_j + c_y \,\delta y_j \textrm{ .}
\end{equation}
Once the best-fit value  $m'_j=c_0 + c_x \,\delta x_j + c_y\,\delta y_j$
had been evaluated for each frame, the corrected light curves were extracted and normalized 
to zero by evaluating $m'_j-m_j$.
\end{enumerate}

Both steps 1) and 2) were applied only when the RMS had decreased. The
resulting median RMS of the PSF-ZP + dith-ZP  correction is plotted in
the right panel of Fig. \ref{cmd} as green circles.  
The small black points in the same plot show the RMS
of each single light curve of cluster members after  our PSF-ZP and dith-ZP  corrections 
have been applied.
Their noise level approaches the theoretical limit in every magnitude bin,
demonstrating that the algorithms we applied were effective.

\section{Light curve analysis}

\subsection{Search for transit-like events}
\label{sec:transits}

\begin{figure*}
\centering \includegraphics[height=14cm]{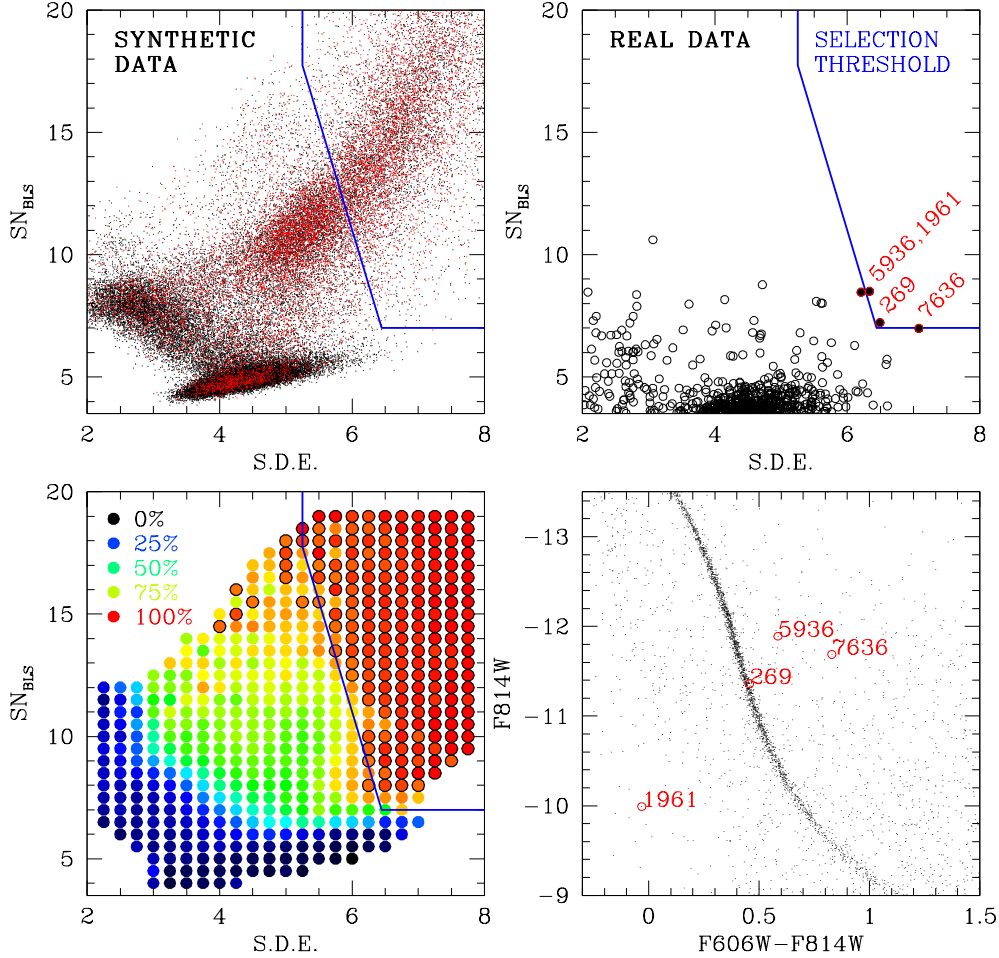}
\caption{ \emph{Upper left:} distribution of SDE and
  $\textrm{SN}_\mathrm{BLS}$ from the BLS analysis of 450,000
  artificially injected transits in  synthetic light curves. ``Recovered''
  transits are plotted as red dots (see text).
  \emph{Lower left:} same as above, where the parameter space has been
  divided into cells and color-coded as a function of the fraction $f$ of
  transits successfully recovered. Cells with $f>90\%$ (that is, with an incidence of false positives smaller than 10\%)
  are highlighted with a black border. The blue line correponds to the
  threshold defined in Eq. (\ref{crit}).  \emph{Upper right:} distribution of SDE and
  $\textrm{SN}_\mathrm{BLS}$ from the BLS analysis of the full sample
  of 5,078 real light curves.  Four low-significance candidates are
  labeled.  \emph{Lower right:} location of the four
  low-significance candidates on the CMD.  }
\label{bls}
\end{figure*}

We searched for transits in the full set of 5,078 light curves (that is,
including  field stars) by applying the box-fitting least-square
algorithm (BLS, \citealt{kovacs2002}).  For each curve, BLS was applied to
search for periodic dips of duration $\Delta$  and depth $\delta$ with
10,000 trial periods  between 0.2 and 14 days. The relative transit
duration $q=\Delta/P$ was constrained to the values possible for planetary
transits around low-MS stars ($R_\star=0.08$--1.4$R_\odot$).
 
For each star, three detection diagnostics were calculated:
 the signal residue (SR), the signal
detection efficiency (SDE) associated with the maximum peak in the BLS
periodogram \citep{kovacs2002}, and the  detection S/N defined
as
\begin{equation}
\textrm{S/N}_\mathrm{BLS}=\frac{\delta}{\sigma}\cdot\sqrt{n_\mathrm{t}}\;\textrm{ ,}
\end{equation}
where $\sigma$ is the (unbinned) photometric noise and $n_\mathrm{t}$ the
number of data points sampled during transits.

Other more sophisticated detection diagnostics, such as the
``signal-to-pink'' S/N   \citep{pont2006,hartman2008}, are 
robust if correlated noise $\sigma_\mathrm{red}$ 
(``red noise'', \citealt{pont2006}) is present. However, they require knowledge of
$\sigma_\mathrm{red}$ on timescales close to $\Delta$. This is
difficult to evaluate in our data, as transits are expected to be undersampled by
the observing cadence. Transits of a  $P\sim 3$~d planet around a M4V
star are expected to last $R_\star P /(\pi a) \sim 60$ min at most, and only $\sim
30$-$40$ min for later types, while images are sampled every  32-64 min. 
However, the amount of red noise here is
very low, as demonstrated by the similarity of the measured RMS to the
theoretical one. Therefore, we decided to employ both $\textrm{S/N}_\mathrm{BLS}$
and SDE as detection diagnostics.

To set a reliable detection criterium, 2,215
light curves were simulated, that had the same sampling times $t_i$ and noise level as
the real M-dwarfs. A synthetic transit (following the
analytical model of \citealt{mandel2002})  of a 1 $R_\mathrm{jup}$
planet was injected into each curve, with a random uniform distribution in  $P$ and
$\sin i$ where $P$ was bounded to the range 1--5 d, while $\sin i$ was
constrained  to allow transits. The process was iterated 200 times for
a total of 450,000  injections. We then tried to recover the transits
with BLS, by setting the same  parameters used for the real search. A
planet was defined as ``recovered'' if at least two transits had been sampled,
and if the estimated orbital period (or a low-order harmonics:  2:1,
3:1, 3:2) matched the injected one. The distributions of the
``injected'' and ``recovered'' transits in the SDE vs. $\textrm{S/N}_\mathrm{BLS}$ 
plane are plotted as black and red points
in the upper left panel of Fig.~\ref{bls}.  By dividing the parameter
space into cells and evaluating the fraction of  ``recovered'' over
``injected'' transits (Fig. \ref{bls}, lower left panel), we obtained
an estimate of the expected fraction $f$ of real transits
successfully detected  (``true positives'').
We defined detection criteria that guarantee a fraction of false
positives that is smaller than 10\% (that is,
$f\geq 90\%$) where 
\begin{equation}\label{crit}
\left\{ \begin{array}{l}  \textrm{SDE} \geq 5.25
  \\ \textrm{S/N}_\mathrm{BLS} \geq 65 -
  9\cdot\textrm{SDE}\\ \textrm{S/N}_\mathrm{BLS} \geq 7 \\
\end{array} \right .
\end{equation}
as indicated by the blue line in Fig. \ref{bls}.  With this choice, the fraction
of false negatives (that is, real planets discarded by selection
criteria) is  about 55\%, and gets larger for planets smaller than
Jupiter.  This is unavoidable if one wishes to keep the  fraction of false positives 
as low as possible.

The position of all the real sources in the (SDE,
$\textrm{SN}_\mathrm{BLS}$) plane is shown in the upper-right panel of
Fig. \ref{bls}. Only four stars among the full sample  meet the
criteria set in Eq. (\ref{crit}) or get very close to the
threshold. These ``borderline'' targets (ID\#0269, 1961, 5936, and
7637) were inspected and cross-checked individually.

\emph{ID\#269} light curve is crippled by a CCD bad pixel falling just
under the star  in one of the dithering positions. 

\emph{ID\#1961} is contaminated by brighter surrounding stars, and its flux drops off
significantly in one fourth of the images. The BLS signal is
probably spurious.

\emph{ID\#5936} seems accurately measured by the reduction pipeline, though
it lies extremely close to a saturated star. Its light curve should be
treated with caution.  The parameters of the detected signal ($P\sim
2.1$ d, $\delta=0.08$ mag, $q=0.025$) would  be compatible with a
$\sim 1$ $R_\mathrm{jup}$ transiting body with zero impact parameter,
or with a grazing eclipsing binary.  We classified ID\#5936 as a
cluster M-dwarf member by analyzing its proper motion (with $M_\star =
0.34 M_\odot$, $R_\star = 0.32 R_\odot$ from its color), though its
position in the CMD diagram  is offset from the MS by 1.4 mag in
$m_\mathrm{F814W}$ and 0.2 mag in color (Fig \ref{bls}, lower right
panel). This cannot be due to binarity alone, and maybe the presence
of a bright contaminant or the departure from the normal evolution of
the companion play a role. There is also a non-negligible probability
that ID\#5936 is a field star having a proper motion 
compatible with the common motion of the cluster.  
It it worth noting that at least one data
point with a similar decrease in brightness (0.08 mag, that is $\sim 4\sigma$) 
fell outside the expected transit windows fitted by BLS, and that the
estimated duration ($\Delta=76$ min) is way larger than expected. 
We do not consider ID\#5936 to be a convincing planetary candidate. 
Its coordinates are listed in Table \ref{vartab} for possible further studies.

\emph{ID\#7636} was rejected because it fell over a bad column on
frames corresponding  to ``transits''. 

In summary, we did not detect transits in our light curves, at least
with an acceptable degree of statistical significance. We discuss the
significance of this null detection in Section \ref{sec:compl}.

\subsection{Search for variable stars}
\label{sec:variables}

\begin{figure*}
\centering \includegraphics[width=17.8cm,trim=0 20 0 0,clip=true]{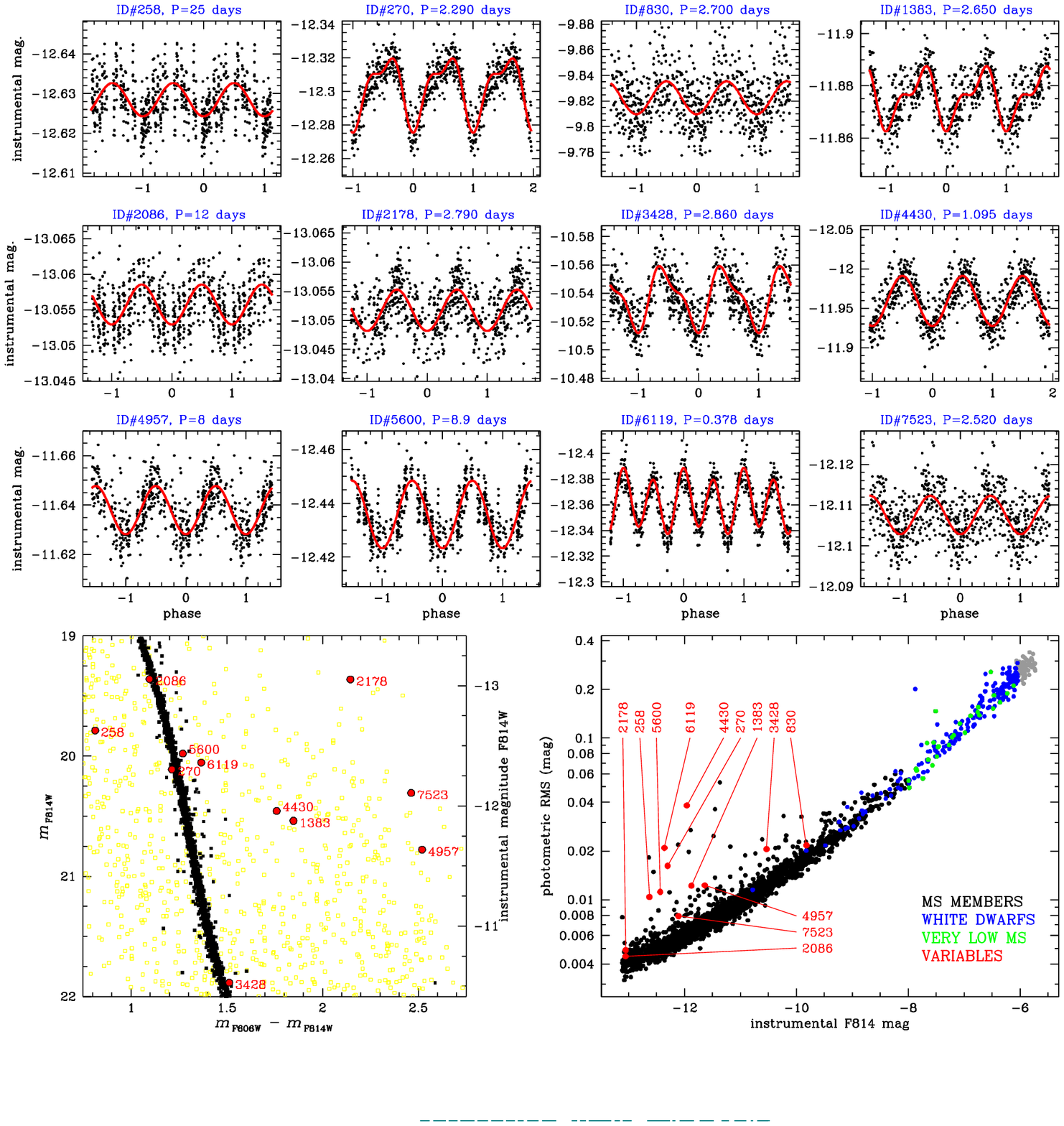}
\caption{\emph{Top panels:} light curves of the variable stars found
 (first twelve entries in Table \ref{vartab}), folded around the best-fit period.
  \emph{Bottom left panel:} position of the variables (red IDs) in the
  ($m_\mathrm{F606W}-m_\mathrm{F814W}$,$m_\mathrm{F814W}$)
  color-magnitude diagram. \emph{Bottom right panel:} photometric RMS
  of the variables (red IDs) compared with all the analyzed light curves.}
\label{var}
\end{figure*}

We performed a search for variable stars in our full database of 5,078
light curves corrected for systematic errors.  First, the
coefficient of spectral correlation \citep{ferraz1981} was calculated
for each light curve. Following the method described in
\citet{demarchi2007,demarchi2010}, we obtained a sample of 13
suspected variable stars (Table \ref{vartab}).  All these candidates,
based on their proper motions and position in the CMD, were identified
as field stars with high confidence.  To classify these objects, 
a least squares iterative sine-wave search was applied
\citep{vanicek1971}. 

Most of our candidates show a single harmonic sinusoidal shape and
short periods ($P\le 9$ d), namely ID\#830, ID\#7523, ID\#2178, ID\#5600, and
ID\#4957.  Without other elements,  it was impossible to derive an
unambiguous classification for these variables.  We suspect that these
stars are most probably field BY Draconis variables, i.e. spotted and
rotating KM dwarfs. This tentative classification is supported
by their very red colors
($m_\mathrm{F606W}-m_\mathrm{F814W}=1.27$-2.65).

For four stars, a good best-fit can be obtained using  two
harmonics. The second harmonic in the light curves of ID\#3428,
ID\#1383, ID\#4430, and ID\#270 could  indicate the presence of spots on the
surfaces and confirm the BY Dra-type classification, while in the
light curve ID\#6119 the second harmonic reveals the profile of a W
UMa contact eclipsing binary system.   Two stars   (ID\#2086 and
ID\#258) show clearly orbit-to-orbit variability but the time coverage
was too short to infer reliable values for their periods: we classified
them as generic ``long period variables''. Finally, the fluctuations
in the light curve ID\#1882 are too small to allow us to confirm its
nature as a variable star, hence we discarded it from our analysis. 
The summary classification of the entire
sample is reported in Table \ref{vartab}, along with the best
candidate transit found by BLS and discussed in Section \ref{sec:transits}.

\linespread{1.2}
\begin{table*}
\caption{Parameters for the variable stars found.}
\label{vartab}
\centering
\scalebox{0.95}{
\begin{tabular}{rcccccccccl}
\hline\hline
ID\# & RA (2000.0) & DEC (2000.0)& $m_\mathrm{F814W}$ & $m_\mathrm{F606W}-m_\mathrm{F814W}$ & p.m.($\alpha$) & p.m.($\delta$) & $P$ & $\Delta m$ & $N_\mathrm{harm}$ & notes\\ 
 & h:m:s & d:m:s &  &   & (pixels) & (pixels) & (days) & (mag) &  & \\ \hline
3428&17:41:06.078&$-$53:45:47.62&21.884&1.513&$-$1.2192&0.4788&2.860&0.047&2&double-wave BY Dra\\
1383&17:41:11.653&$-$53:45:02.54&20.538&1.848&$-$1.3037&0.2494&2.650&0.025&2&double-wave BY Dra\\
 270&17:41:14.751&$-$53:45:04.94&20.110&1.213&$-$1.0478&0.7856&2.290&0.044&2&double-wave BY Dra\\
6119&17:40:58.818&$-$53:45:41.10&20.054&1.367&$-$1.4436&0.5940&0.378&0.051&2&W UMa\\
4430&17:41:03.155&$-$53:44:49.11&20.456&1.761&$-$1.2356&0.4401&1.095&0.069&2&BY Dra?\\
 830&17:41:13.183&$-$53:45:37.72&22.598&2.654&$-$1.3962&0.6337&2.700&0.026&1&BY Dra?\\
7523&17:40:55.370&$-$53:45:27.03&20.306&2.463&\,\,1.0304&0.5997&2.520&0.009&1&BY Dra? (weak)\\
2178&17:41:09.636&$-$53:43:17.61&19.362&2.146&$-$1.6238&0.5057&2.790&0.007&1&BY Dra? (weak)\\
5600&17:41:00.192&$-$53:42:46.28&19.979&1.271&$-$1.5276&0.3267&$\approx$ 8.9 &0.025&1& BY Dra?\\
4957&17:41:01.787&$-$53:43:31.37&20.778&2.520&$-$1.6460&0.3394&$\approx$ 8&0.019&1&BY Dra?\\
2086&17:41:09.888&$-$53:43:10.59&19.359&1.097&$-$1.5737&0.3106&$\approx$ 12&0.006&1&long period\\
 258&17:41:14.789&$-$53:44:58.36&19.786&0.813&$-$1.7434&0.3614&$\approx$ 25 &0.008&1&long period\\
1882&17:41:10.379&$-$53:44:52.05&19.430&0.792&$-$1.0721&0.7363&0.920&0.005&1 & spurious candidate?\\ \hline
5936&17:40:59.316&$-$53:44:06.51&20.524&1.493&$\phantom{-}$0.0181&0.0055&2.120&0.080&-- & grazing binary?\\ \hline
\end{tabular} }
\tablefoot{The columns give: the ID number of the star, the right ascension $\alpha$ and declination $\delta$ 
at epoch 2000.0, the calibrated magnitude $m$(F814W) and color $m$(F814W)$-m$(F606W), the proper motion ($\alpha$,$\delta$) 
in ACS-WFC pixels relative to the cluster, the period found in days, the amplitude found in magnitude, and
the number of harmonics employed in the fit and a tentative interpretation.}
\end{table*}
\linespread{1.0}

\section{Completeness and significance}
\label{sec:compl}

\begin{figure*}
\centering
\includegraphics[height=17cm, angle=-90]{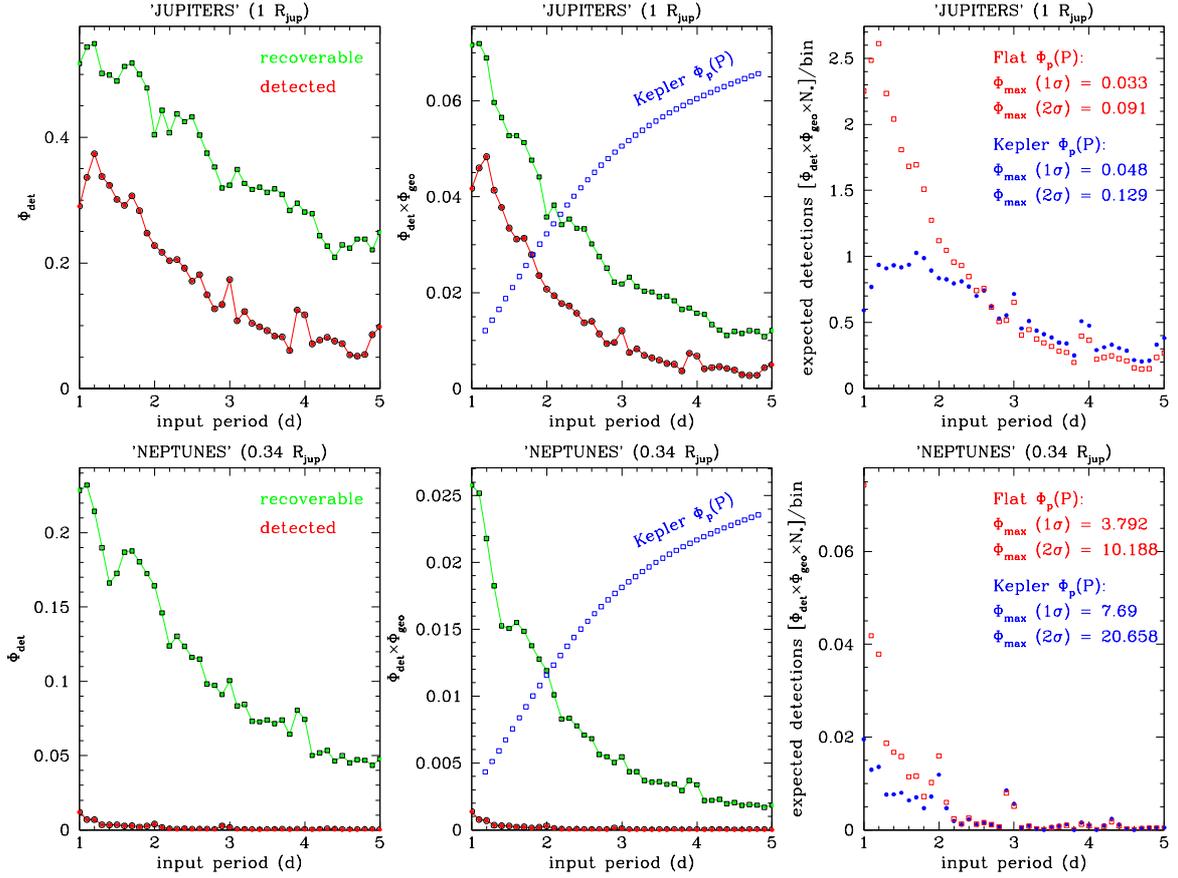}
\caption{Completeness tests for cluster stars based on artificially injected transits,
for 1 $R_\mathrm{jup}$ planets (upper row) and 0.336 $R_\mathrm{jup}$ planets (lower row).
\emph{Left panels:} detection efficiency $\Phi_\mathrm{det}$ as a
function of the input period, for planets potentially recoverable (green symbols)
and for those planets effectively detected by the criterion in Eq. (\ref{crit}) (red symbols).
\emph{Middle panels:} same as above, but $\Phi_\mathrm{det}$ is convolved with 
the geometrical probability $\Phi_\mathrm{geo}$  for a planet to transit. The \citet{howard2011} 
$\Phi_\mathrm{p}(P)$ period distribution function is plotted in blue symbols (arbitrary
normalization).
\emph{Right panels:} number of expected transit detections per period bin, assuming 
one planet per star within $1<P<5$ d. $\Phi_\mathrm{p}(P)$ is assumed to be flat
(red symbols) or as modeled by \citet{howard2011} (blue symbols). 1- and 2-$\sigma$ upper limits 
for the planet occurence $\Phi_\mathrm{p,\max}$ are shown.
}
\label{compl}
\end{figure*}

The significance of our null detection of transits was assessed by
considering only  the 2,215 cluster-member M dwarfs for which we 
derived reliable estimates  of $R_\star$ and $M_\star$ from their
position in the CMD. 

For a planet of given radius $R_\mathrm{p}$ and orbital period $P$,
the number of expected planet detections is given by
\begin{equation}
N_\mathrm{p} = N_\star\iint \left (
\Phi_\mathrm{p}(P,R_\mathrm{p})\cdot\Phi_\mathrm{geo}\cdot\Phi_\mathrm{det}\right
)\,\mathrm{d}P\,\mathrm{d}R_\mathrm{p} \;\textrm{ ,}
\end{equation}
where $\Phi_\mathrm{p}$ is the fraction of stars with a planet,
$\Phi_\mathrm{geo}$ is the geometric \emph{a priori} probability for that system to be
aligned such that a transit occurs, $\Phi_\mathrm{det}$ is the
probability for that transit to be detected by our pipeline, and $N_\star$
is the number of target stars.  As in our case we found that
$N_\mathrm{p}=0$, we wished to estimate an upper limit to
$\Phi_\mathrm{p}$, at least in the $(P,R_\mathrm{p}$) range for which the
efficiency of our search $\Phi_\mathrm{det}$ is not
negligible. $\Phi_\mathrm{det}$ is expected to depend on the
transit depth $(R_\mathrm{p}/R_\star)^2$, the duration $\Delta$, and
the orbital period $P$.

For simplicity, our analysis was limited to two values of planetary
radii:   ``Jupiter'' planets (1 $R_\mathrm{jup}$) and ``Neptune''
planets (0.338 $R_\mathrm{jup}$).  To estimate $\Phi_\mathrm{det}$, we
ran simulations in a way similar to what has been done to set the
detection threshold (Section \ref{sec:transits}). In each set of 2,215
simulated  light curves, a synthetic transit was injected into each curve,
with random uniform distributions of both  $P$ and $\sin i$ ($1\leq P \leq
5$ d, $\sin i$ was constrained  to allow transits). The process was
iterated 200 times, for a total of 450,000  injections. We then tried
to recover transits with BLS, by setting the same parameters and detection
criterium defined in Eq. (\ref{crit}) and adopted for the real search.   To derive
$\Phi_\mathrm{det}$, the ratio of the detected to injected transits
was evaluated for each 0.1-day bin of period $P$.  
The resulting distribution is plotted as a red line in the left panels of 
Fig. \ref{compl}, as a function of the injected orbital period $P_\mathrm{in}$.
For comparison, the ratio of the ``recoverable'' transits 
(i.e., with at least two transits sampled and $P/P_\mathrm{in}=$1:1, 3:2, 2:1) 
to the injected transits is plotted with a green line on the same panels.
As expected,
$\Phi_\mathrm{det}$ is a decreasing function of $P$, with minor
features at integer and semi-integer values of $P$ owing to phasing
effects. We note that for ``Neptunes'', $\Phi_\mathrm{det}$ is extremely
low (0.005--0.01) even for short periods ($P\sim 1$--2 d). This is a
consequence of inefficient sampling, which makes the significance of
neptunian transits very weak: 3--$4\sigma$  even for the most
favourable case.

The geometric factor $\Phi_\mathrm{geo}$ was calculated for each
injected  transit as $(R_\mathrm{p}+R_\star)/a$ ($a$ is the semimajor
axis), and  then convolved with $\Phi_\mathrm{det}$ to obtain the
probability of detecting transits on a star which is known  to host a
planet on a random orbit, as a function of its period (Fig. \ref{compl},
middle panels). 

We parametrized the ``planet occurrence''   following the analysis of
\citet{howard2011} for the distribution of 1,235  planetary candidates
detected by Kepler.  In this case, $\Phi_\mathrm{p}(P)$ was assumed to be a power law
modified with an exponential cut-off at period $P_\mathrm{cut}$
\begin{equation}
\frac{\mathrm{d}\Phi_\mathrm{p}(P)}{\mathrm{d}\log P} = k\cdot P^{\,\beta}
\left ( 1-\mathrm{e}^{-(P/P_\mathrm{cut})^\gamma}  \right )\;\textrm{ .}
\end{equation}
From \citet{howard2011} we adopted the parameters: $k=0.0025$, $\beta=0.37$,
$P_\mathrm{cut}=1.7$ days, $\gamma = 4.1$  for ``Jupiters'' 
($8R_\oplus<R_\mathrm{p}<32R_\oplus$), and $k=0.002$, $\beta=0.79$, $P_\mathrm{cut}=2.2$
days, and $\gamma = 4.0$  for ``Neptunes'' 
($4R_\oplus<R_\mathrm{p}<8R_\oplus$). $\Phi_\mathrm{p}(P)$ is plotted, with an arbitrary
normalization, as blue squares in the middle panels of
Fig. \ref{compl}. 

We first normalized $\Phi_\mathrm{p}(P)$ by imposing $\sum_1^5
\Phi_\mathrm{p}(P)=1$, that is assuming one planet with 1 d $<P<5$ d per
star. The total number of expected detections
$N_\mathrm{exp}(\sum\Phi_\mathrm{p}=1)$ within each bin over the range
$1<P<5$ d is
($\Phi_\mathrm{det}\cdot\Phi_\mathrm{geo}\cdot\Phi_\mathrm{p})\,
N_\star$. By summing over the range $1<P<5$ d, we obtained
$N_\mathrm{exp}=23.8$ expected detections of ``Jupiters'' and 0.14 of ``Neptunes'' (right
panels of Fig. \ref{compl}, blue  symbols). For a flat
$\Phi_\mathrm{p}(P)$ distribution, $N_\mathrm{exp}$ 
is larger at 42.3 and 0.45, respectively.

The upper limit $\Phi_\mathrm{p,max}$ to $\sum_1^5 \Phi_\mathrm{p}(P)$
suggested by our null detection can be evaluated by simple binomial
statistics, normalizing $\Phi_\mathrm{p}$ in order to get  a
68.27\% (1$\sigma$) or 95.44\% (2$\sigma$) probability of zero detections. 
We estimated for Jupiters that $\Phi_\mathrm{p,max}(1\sigma)=4.8\%$ and
$\Phi_\mathrm{p,max}(2\sigma)=12.9\%$ assuming the \citet{howard2011}
$\Phi_\mathrm{p}$, and $\Phi_\mathrm{p,max}(1\sigma)=3.3\%$,
$\Phi_\mathrm{p,max}(2\sigma)=9.1\%$  assuming a flat
$\Phi_\mathrm{p}$. As expected, $\Phi_\mathrm{p,max}$ is well above
unity for the ``Neptune'' sample, leaving this  planetary population
essentially unconstrained by our data (Fig. \ref{compl}, lower right
panel).

\section{Discussion and conclusions}

We have performed a search for planetary transits and variability among
5,078 stars imaged in one of the deepest ACS fields ever observed, 
which had been originally acquired to probe the bottom of the main sequence of the metal-poor globular
cluster NGC 6397. The sample includes 2,215  M0-M9 dwarfs of secure
membership. Though these data were not optimized for such a study,
this is the largest homogeneous sample of M dwarfs ever searched for
variability. 

Instrumental drifts and systematic errors caused by
dithering required a careful empirical correction, described in
Section \ref{sec:zp}. We developed and implemented algorithms that allowed us
to approach the theoretical noise limit across the whole magnitude range
$19 \lesssim m_\mathrm{F814W}\lesssim 26$. 
The brightest cluster members (M0V) were measured with an average scatter of 
0.003-0.004 mag over a time span of 28 days, illustrating the 
power of our decorrelating techniques and the feasibility of
transit searches in the low main sequence of GCs.

We found no valid planetary transit above the significance threshold
that we set from simulations. Considering only cluster stars, whose
physical parameters can be reliabily infered, this null detection sets
an upper limit to the fraction of stars hosting a $P<5$ d
Jupiter-sized planet to about $\Phi_p =4.8$\% at 1-$\sigma$ confidence 
and 12.9\% at 2-$\sigma$, assuming  the planetary radius distribution
derived by \citet{howard2011} from Kepler data. 
In other words, only 0.13 detections are expected assuming
that the underlying planetary population is similar to that
studied by Kepler. Most studies based on RVs also
hypotesized $\Phi_p <1$\% for short-period, Jupiter-sized planets
around solar-type stars \citep{marcy2005}.  Furthermore,
$\Phi_p$ is expected to be a very steep function of the stellar
metallicity \citep{fischervalenti2005}. Therefore, we are unable to make 
any firm conclusion about the occurrence of giant planets in NGC 6397.

As demonstrated in Section
\ref{sec:transits}, our data set is not sensitive enough to
Neptune-sized planets to draw any  conclusion about their occurrence,
though a much higher $\Phi_p$ is expected for M dwarfs by
\citet{howard2011} and \citet{lovis2009}, among others.  This was due
to an inefficient sampling of the time series available from archive
material, which translated into
poor phase coverage and severe undersampling of transit-like events,
whose duration is expected to be on the same order as the effective
cadence.

Twelve new variable stars have been identified in the NGC 6397 field
(Table \ref{vartab}).  Most of these can be classified as BY Draconis
variables, that is, spotted rotating KM dwarfs. Interestingly, no
variable has been detected among the 2,430 cluster members.  Hundreds
of member early-M dwarfs ($m_\mathrm{F814W}<21$, Fig. \ref{cmd})
follow the expected noise on timescales of up to 28 days, though they
were measured with a 0.003-0.006 mag precision. The lack of
eclipsing binaries is unsurprising. 
The number of expected detections can be
estimated by  scaling down the number of EBs detected by \citet{albrow2001} for 47 Tuc.
If one takes into account the smaller fraction of binaries in NGC 6397 
($<3$\%, even at radii smaller than the half-mass
radius, \citealt{milone2011}) and the smaller number of targets (2,215 vs. 46,422),
we would expect much less than one EB. 
On the other hand, the lack of BY Dra 
variables is more puzzling, as one would expect 4-5 such detections, 
considering the number of monitored cluster stars in this paper.
We note that our stars are cooler than the 
\citet{albrow2001} sample, and we should expect longer photometric periods. 
Our search is insensitive to periods $\gtrsim 28$ days, so this could be a 
possible explanation.

We evaluated an upper
limit to the photometric jitter $\sigma_\mathrm{jit}$ of the brightest
members (M0V) by subtracting the contribution of the expected noise
$\sigma_\mathrm{exp}$ to the measured scatter $\sigma_\mathrm{obs}$
(that is, assuming $\sigma_\mathrm{jit}^2 = \sigma_\mathrm{obs}^2 -
\sigma_\mathrm{exp}^2$).  The fraction of stars $f$ with
$\sigma_\mathrm{jit}>2$ mmag is $f\lesssim 2\%$.  This value should be
compared with the results  found by \citet{ciardi2011} examining the
first quarter of Kepler photometry on 2,182 field M dwarfs:  these
data cover an interval of 33 days with an average cadence of 30 minutes,
which is
quite similar to our cadence and time scale.  They found a fraction
$f\simeq 20\%$ of stars with $\sigma > 2$ mmag,  that is at least an
order of magnitude larger fraction than that we measured in NGC 6397.  The low
MS of this cluster is extremely stable and therefore worth targeting
using more optimized observations.  The James Webb Space Telescope 
Near-Infrared Camera (NIRCam), for instance,
would be able to probe the bottom of the MS of NGC 6397
($m_\mathrm{F814W}\sim24$) with a photometric precision better than
0.01 mag in a single 600 s exposure, without any of the
coverage/sampling issues mentioned above. 

Most of the analysis techniques presented in this paper  can also be
applied (with little or no modification) to other existing  ACS/WFC3
time series of rich stellar fields. This is the case for the metal-rich
globular cluster 47 Tucanae, which has been imaged with ACS and WFC3  
over a longer time-frame. Such a search for
transits  will allow us to complement the results of \citet{gilliland2000}
in a  different range of spectral types and planetary masses.

\begin{acknowledgements}
This work was partially supported by PRIN INAF 2008 "Environmental effects in the formation and evolution of extrasolar planetary
system". V.N. acknowledges support by STScI grant DDRF D0001.82432.
We thank Ennio Poretti for helping us to identify the
variable stars.
We thank Aaron Dotter for providing us with the isochrones used
in \citet{richer2008}  and presented in \citet{dotter2007}.  Some
tasks of our data analysis have been carried out with the VARTOOLS
code \citep{hartman2008}. We thank Ron Gilliland
for his useful comments and suggestions.

\end{acknowledgements}

\medskip

\emph{NOTE to the astro-ph version.} The referee asked us if
it is possible that the proper motions have systematic 
problems for the variable stars.

\begin{figure}
\centering \includegraphics[width=\columnwidth]{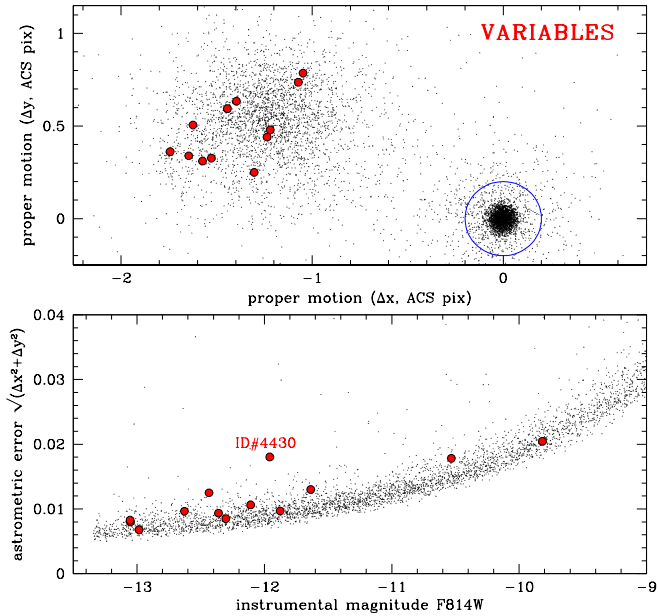}
\caption{(\textbf{not in the published version}).
\emph{Upper panel:} relative proper motions for all individual
stars (black dots) with the twelve identified variables plotted with
red circles. The blue circle marks the average proper motion of NGC 6397.
\emph{Lower panel:} mean astrometric error as a function of instrumental
magnitude, color-coded as above. 
See text in the final note for details. In both panels, units are
ACS physical pixels.}
\label{variables}
\end{figure}

We doubt that variability could have biased our proper
motions. In fact, our variables have proper motions statistically
consistent with those measured for non-variable field stars (see 
Fig.~\ref{variables},
 upper panel, red points). Also the astrometric
error is consistent with that found on stars of similar magnitude,
with the only exception of ID\#4430 (Fig.~\ref{variables}, lower 
panel). ID\#4430 is the only variable contaminated by
a nearby star; however, it does not lie close to the MS in the CMD.

\bibliographystyle{aa}
\bibliography{biblio}

\begin{thebibliography}{50}
\expandafter\ifx\csname natexlab\endcsname\relax\def\natexlab#1{#1}\fi

\bibitem[{{Albrow} {et~al.}(2001){Albrow}, {Gilliland}, {Brown}, {Edmonds},
  {Guhathakurta}, \& {Sarajedini}}]{albrow2001}
{Albrow}, M.~D., {Gilliland}, R.~L., {Brown}, T.~M., {et~al.} 2001, \apj, 559,
  1060

\bibitem[{{Anderson} \& {Bedin}(2010)}]{andersonbedin2010}
{Anderson}, J. \& {Bedin}, L.~R. 2010, \pasp, 122, 1035

\bibitem[{{Anderson} \& {King}(2000)}]{andersonking2000}
{Anderson}, J. \& {King}, I.~R. 2000, \pasp, 112, 1360

\bibitem[{{Anderson} \& {King}(2006)}]{andersonking2006}
{Anderson}, J. \& {King}, I.~R. 2006, {PSFs, Photometry, and Astronomy for the
  ACS/WFC}, Tech. rep.

\bibitem[{{Anderson} {et~al.}(2008){Anderson}, {King}, {Richer}, {Fahlman},
  {Hansen}, {Hurley}, {Kalirai}, {Rich}, \& {Stetson}}]{anderson2008}
{Anderson}, J., {King}, I.~R., {Richer}, H.~B., {et~al.} 2008, \aj, 135, 2114

\bibitem[{{Borucki} {et~al.}(2010){Borucki}, {Koch}, {Basri}, {Batalha},
  {Brown}, {Caldwell}, {Caldwell}, {Christensen-Dalsgaard}, {Cochran},
  {DeVore}, {Dunham}, {Dupree}, {Gautier}, {Geary}, {Gilliland}, {Gould},
  {Howell}, {Jenkins}, {Kondo}, {Latham}, {Marcy}, {Meibom}, {Kjeldsen},
  {Lissauer}, {Monet}, {Morrison}, {Sasselov}, {Tarter}, {Boss}, {Brownlee},
  {Owen}, {Buzasi}, {Charbonneau}, {Doyle}, {Fortney}, {Ford}, {Holman},
  {Seager}, {Steffen}, {Welsh}, {Rowe}, {Anderson}, {Buchhave}, {Ciardi},
  {Walkowicz}, {Sherry}, {Horch}, {Isaacson}, {Everett}, {Fischer}, {Torres},
  {Johnson}, {Endl}, {MacQueen}, {Bryson}, {Dotson}, {Haas}, {Kolodziejczak},
  {Van Cleve}, {Chandrasekaran}, {Twicken}, {Quintana}, {Clarke}, {Allen},
  {Li}, {Wu}, {Tenenbaum}, {Verner}, {Bruhweiler}, {Barnes}, \&
  {Prsa}}]{borucki2010}
{Borucki}, W.~J., {Koch}, D., {Basri}, G., {et~al.} 2010, Science, 327, 977

\bibitem[{{Broeg} {et~al.}(2005){Broeg}, {Fern{\'a}ndez}, \&
  {Neuh{\"a}user}}]{broeg2005}
{Broeg}, C., {Fern{\'a}ndez}, M., \& {Neuh{\"a}user}, R. 2005, Astronomische
  Nachrichten, 326, 134

\bibitem[{{Ciardi} {et~al.}(2011){Ciardi}, {von Braun}, {Bryden}, {van Eyken},
  {Howell}, {Kane}, {Plavchan}, {Ram{\'{\i}}rez}, \& {Stauffer}}]{ciardi2011}
{Ciardi}, D.~R., {von Braun}, K., {Bryden}, G., {et~al.} 2011, \aj, 141, 108

\bibitem[{{de Marchi} {et~al.}(2010){de Marchi}, {Poretti}, {Montalto},
  {Desidera}, \& {Piotto}}]{demarchi2010}
{de Marchi}, F., {Poretti}, E., {Montalto}, M., {Desidera}, S., \& {Piotto}, G.
  2010, \aap, 509, A17

\bibitem[{{de Marchi} {et~al.}(2007){de Marchi}, {Poretti}, {Montalto},
  {Piotto}, {Desidera}, {Bedin}, {Claudi}, {Arellano Ferro}, {Bruntt}, \&
  {Stetson}}]{demarchi2007}
{de Marchi}, F., {Poretti}, E., {Montalto}, M., {et~al.} 2007, \aap, 471, 515

\bibitem[{{Dotter} {et~al.}(2007){Dotter}, {Chaboyer}, {Jevremovi{\'c}},
  {Baron}, {Ferguson}, {Sarajedini}, \& {Anderson}}]{dotter2007}
{Dotter}, A., {Chaboyer}, B., {Jevremovi{\'c}}, D., {et~al.} 2007, \aj, 134,
  376

\bibitem[{{Ferraz-Mello}(1981)}]{ferraz1981}
{Ferraz-Mello}, S. 1981, \aj, 86, 619

\bibitem[{{Fischer} \& {Valenti}(2005)}]{fischervalenti2005}
{Fischer}, D.~A. \& {Valenti}, J. 2005, \apj, 622, 1102

\bibitem[{{Fregeau} {et~al.}(2006){Fregeau}, {Chatterjee}, \&
  {Rasio}}]{fregeau2006}
{Fregeau}, J.~M., {Chatterjee}, S., \& {Rasio}, F.~A. 2006, \apj, 640, 1086

\bibitem[{{Gilliland} {et~al.}(2000){Gilliland}, {Brown}, {Guhathakurta},
  {Sarajedini}, {Milone}, {Albrow}, {Baliber}, {Bruntt}, {Burrows},
  {Charbonneau}, {Choi}, {Cochran}, {Edmonds}, {Frandsen}, {Howell}, {Lin},
  {Marcy}, {Mayor}, {Naef}, {Sigurdsson}, {Stagg}, {Vandenberg}, {Vogt}, \&
  {Williams}}]{gilliland2000}
{Gilliland}, R.~L., {Brown}, T.~M., {Guhathakurta}, P., {et~al.} 2000, \apjl,
  545, L47

\bibitem[{{Gratton} {et~al.}(2003){Gratton}, {Bragaglia}, {Carretta},
  {Clementini}, {Desidera}, {Grundahl}, \& {Lucatello}}]{gratton2003}
{Gratton}, R.~G., {Bragaglia}, A., {Carretta}, E., {et~al.} 2003, \aap, 408,
  529

\bibitem[{{Hansen} {et~al.}(2007){Hansen}, {Anderson}, {Brewer}, {Dotter},
  {Fahlman}, {Hurley}, {Kalirai}, {King}, {Reitzel}, {Richer}, {Rich}, {Shara},
  \& {Stetson}}]{hansen2007}
{Hansen}, B.~M.~S., {Anderson}, J., {Brewer}, J., {et~al.} 2007, \apj, 671, 380

\bibitem[{{Hartman} {et~al.}(2008){Hartman}, {Gaudi}, {Holman}, {McLeod},
  {Stanek}, {Barranco}, {Pinsonneault}, \& {Kalirai}}]{hartman2008}
{Hartman}, J.~D., {Gaudi}, B.~S., {Holman}, M.~J., {et~al.} 2008, \apj, 675,
  1254

\bibitem[{{Hartman} {et~al.}(2009){Hartman}, {Gaudi}, {Holman}, {McLeod},
  {Stanek}, {Barranco}, {Pinsonneault}, {Meibom}, \& {Kalirai}}]{hartman2009}
{Hartman}, J.~D., {Gaudi}, B.~S., {Holman}, M.~J., {et~al.} 2009, \apj, 695,
  336

\bibitem[{{Howard} {et~al.}(2011){Howard}, {Marcy}, {Bryson}, {Jenkins},
  {Rowe}, {Batalha}, {Borucki}, {Koch}, {Dunham}, {Gautier}, {Van Cleve},
  {Cochran}, {Latham}, {Lissauer}, {Torres}, {Brown}, {Gilliland}, {Buchhave},
  {Caldwell}, {Christensen-Dalsgaard}, {Ciardi}, {Fressin}, {Haas}, {Howell},
  {Kjeldsen}, {Seager}, {Rogers}, {Sasselov}, {Steffen}, {Basri},
  {Charbonneau}, {Christiansen}, {Clarke}, {Dupree}, {Fabrycky}, {Fischer},
  {Ford}, {Fortney}, {Tarter}, {Girouard}, {Holman}, {Johnson}, {Klaus},
  {Machalek}, {Moorhead}, {Morehead}, {Ragozzine}, {Tenenbaum}, {Twicken},
  {Quinn}, {Isaacson}, {Shporer}, {Lucas}, {Walkowicz}, {Welsh}, {Boss},
  {Devore}, {Gould}, {Smith}, {Morris}, {Prsa}, \& {Morton}}]{howard2011}
{Howard}, A.~W., {Marcy}, G.~W., {Bryson}, S.~T., {et~al.} 2011, ArXiv e-prints

\bibitem[{{Johnson} {et~al.}(2010){Johnson}, {Aller}, {Howard}, \&
  {Crepp}}]{johnson2010}
{Johnson}, J.~A., {Aller}, K.~M., {Howard}, A.~W., \& {Crepp}, J.~R. 2010,
  \pasp, 122, 905

\bibitem[{{Johnson} \& {Apps}(2009)}]{johnson2009}
{Johnson}, J.~A. \& {Apps}, K. 2009, \apj, 699, 933

\bibitem[{{Kov{\'a}cs} {et~al.}(2002){Kov{\'a}cs}, {Zucker}, \&
  {Mazeh}}]{kovacs2002}
{Kov{\'a}cs}, G., {Zucker}, S., \& {Mazeh}, T. 2002, \aap, 391, 369

\bibitem[{{Lovis} \& {Mayor}(2007)}]{lovismayor2007}
{Lovis}, C. \& {Mayor}, M. 2007, \aap, 472, 657

\bibitem[{{Lovis} {et~al.}(2009){Lovis}, {Mayor}, {Bouchy}, {Pepe}, {Queloz},
  {Udry}, {Benz}, \& {Mordasini}}]{lovis2009}
{Lovis}, C., {Mayor}, M., {Bouchy}, F., {et~al.} 2009, in IAU Symposium, Vol.
  253, IAU Symposium, 502--505

\bibitem[{{Mandel} \& {Agol}(2002)}]{mandel2002}
{Mandel}, K. \& {Agol}, E. 2002, \apjl, 580, L171

\bibitem[{{Marcy} {et~al.}(2005){Marcy}, {Butler}, {Fischer}, {Vogt}, {Wright},
  {Tinney}, \& {Jones}}]{marcy2005}
{Marcy}, G., {Butler}, R.~P., {Fischer}, D., {et~al.} 2005, Progress of
  Theoretical Physics Supplement, 158, 24

\bibitem[{{Milone} {et~al.}(2011){Milone}, {Piotto}, {Bedin}, {Aparicio},
  {Anderson}, {Sarajedini}, {Marino}, {Moretti}, {Davies}, {Chaboyer},
  {Dotter}, {Hempel}, {Marin-Franch}, {Majewski}, {Paust}, {Reid}, {Rosenberg},
  \& {Siegel}}]{milone2011}
{Milone}, A.~P., {Piotto}, G., {Bedin}, L.~R., {et~al.} 2011,
  \texttt{arxiv:1111.0552}

\bibitem[{{Mochejska} {et~al.}(2006){Mochejska}, {Stanek}, {Sasselov},
  {Szentgyorgyi}, {Adams}, {Cooper}, {Foster}, {Hartman}, {Hickox}, {Lai},
  {Westover}, \& {Winn}}]{mochejska2006}
{Mochejska}, B.~J., {Stanek}, K.~Z., {Sasselov}, D.~D., {et~al.} 2006, \aj,
  131, 1090

\bibitem[{{Mochejska} {et~al.}(2005){Mochejska}, {Stanek}, {Sasselov},
  {Szentgyorgyi}, {Bakos}, {Hradecky}, {Devor}, {Marrone}, {Winn}, \&
  {Zaldarriaga}}]{mochejska2005}
{Mochejska}, B.~J., {Stanek}, K.~Z., {Sasselov}, D.~D., {et~al.} 2005, \aj,
  129, 2856

\bibitem[{{Montalto} {et~al.}(2007){Montalto}, {Piotto}, {Desidera}, {de
  Marchi}, {Bruntt}, {Stetson}, {Arellano Ferro}, {Momany}, {Gratton},
  {Poretti}, {Aparicio}, {Barbieri}, {Claudi}, {Grundahl}, \&
  {Rosenberg}}]{montalto2007}
{Montalto}, M., {Piotto}, G., {Desidera}, S., {et~al.} 2007, \aap, 470, 1137

\bibitem[{{Montalto} {et~al.}(2011){Montalto}, {Villanova}, {Koppenhoefer},
  {Piotto}, {Desidera}, {de Marchi}, {Poretti}, {Bedin}, \&
  {Saviane}}]{montalto2011}
{Montalto}, M., {Villanova}, S., {Koppenhoefer}, J., {et~al.} 2011, \aap, 535,
  A39

\bibitem[{{Nascimbeni} {et~al.}(2011){Nascimbeni}, {Piotto}, {Bedin}, \&
  {Damasso}}]{nascimbeni2010}
{Nascimbeni}, V., {Piotto}, G., {Bedin}, L.~R., \& {Damasso}, M. 2011, \aap,
  527, A85+

\bibitem[{{Pont} {et~al.}(2006){Pont}, {Zucker}, \& {Queloz}}]{pont2006}
{Pont}, F., {Zucker}, S., \& {Queloz}, D. 2006, \mnras, 373, 231

\bibitem[{{Richer} {et~al.}(2008){Richer}, {Dotter}, {Hurley}, {Anderson},
  {King}, {Davis}, {Fahlman}, {Hansen}, {Kalirai}, {Paust}, {Rich}, \&
  {Shara}}]{richer2008}
{Richer}, H.~B., {Dotter}, A., {Hurley}, J., {et~al.} 2008, \aj, 135, 2141

\bibitem[{{Santos} {et~al.}(2011){Santos}, {Mayor}, {Bonfils}, {Dumusque},
  {Bouchy}, {Figueira}, {Lovis}, {Melo}, {Pepe}, {Queloz}, {S{\'e}gransan},
  {Sousa}, \& {Udry}}]{santos2011}
{Santos}, N.~C., {Mayor}, M., {Bonfils}, X., {et~al.} 2011, \aap, 526, A112+

\bibitem[{{Sato} {et~al.}(2007){Sato}, {Izumiura}, {Toyota}, {Kambe}, {Takeda},
  {Masuda}, {Omiya}, {Murata}, {Itoh}, {Ando}, {Yoshida}, {Ikoma}, {Kokubo}, \&
  {Ida}}]{sato2007}
{Sato}, B., {Izumiura}, H., {Toyota}, E., {et~al.} 2007, \apj, 661, 527

\bibitem[{{Scalo} {et~al.}(2007){Scalo}, {Kaltenegger}, {Segura}, {Fridlund},
  {Ribas}, {Kulikov}, {Grenfell}, {Rauer}, {Odert}, {Leitzinger}, {Selsis},
  {Khodachenko}, {Eiroa}, {Kasting}, \& {Lammer}}]{scalo2007}
{Scalo}, J., {Kaltenegger}, L., {Segura}, A.~G., {et~al.} 2007, Astrobiology,
  7, 85

\bibitem[{{Schlaufman} \& {Laughlin}(2011)}]{schlaufman2011}
{Schlaufman}, K.~C. \& {Laughlin}, G. 2011, ArXiv e-prints

\bibitem[{{Seager}(2011)}]{seager2011}
{Seager}, S. 2011, {Exoplanets}, ed. {Piper, S.}

\bibitem[{{Spurzem} {et~al.}(2009){Spurzem}, {Giersz}, {Heggie}, \&
  {Lin}}]{spurzem2009}
{Spurzem}, R., {Giersz}, M., {Heggie}, D.~C., \& {Lin}, D.~N.~C. 2009, \apj,
  697, 458

\bibitem[{{Stello} \& {Gilliland}(2009)}]{stello2009}
{Stello}, D. \& {Gilliland}, R.~L. 2009, \apj, 700, 949

\bibitem[{{Thorsett} {et~al.}(1999){Thorsett}, {Arzoumanian}, {Camilo}, \&
  {Lyne}}]{thorsett1999}
{Thorsett}, S.~E., {Arzoumanian}, Z., {Camilo}, F., \& {Lyne}, A.~G. 1999,
  \apj, 523, 763

\bibitem[{{Udry} {et~al.}(2006){Udry}, {Mayor}, {Benz}, {Bertaux}, {Bouchy},
  {Lovis}, {Mordasini}, {Pepe}, {Queloz}, \& {Sivan}}]{udry2006}
{Udry}, S., {Mayor}, M., {Benz}, W., {et~al.} 2006, \aap, 447, 361

\bibitem[{{Udry} \& {Santos}(2007)}]{udry2007}
{Udry}, S. \& {Santos}, N.~C. 2007, \araa, 45, 397

\bibitem[{{van Saders} \& {Gaudi}(2011)}]{vansaders2011}
{van Saders}, J.~L. \& {Gaudi}, B.~S. 2011, \apj, 729, 63

\bibitem[{{Van{\'{\i}}{\v c}ek}(1971)}]{vanicek1971}
{Van{\'{\i}}{\v c}ek}, P. 1971, \apss, 12, 10

\bibitem[{{Weldrake} {et~al.}(2008){Weldrake}, {Sackett}, \&
  {Bridges}}]{weldrake2008}
{Weldrake}, D.~T.~F., {Sackett}, P.~D., \& {Bridges}, T.~J. 2008, \apj, 674,
  1117

\bibitem[{{Weldrake} {et~al.}(2005){Weldrake}, {Sackett}, {Bridges}, \&
  {Freeman}}]{weldrake2005}
{Weldrake}, D.~T.~F., {Sackett}, P.~D., {Bridges}, T.~J., \& {Freeman}, K.~C.
  2005, \apj, 620, 1043

\bibitem[{{Wittenmyer} {et~al.}(2011){Wittenmyer}, {Tinney}, {Butler},
  {O'Toole}, {Jones}, {Carter}, {Bailey}, \& {Horner}}]{wittenmyer2011}
{Wittenmyer}, R.~A., {Tinney}, C.~G., {Butler}, R.~P., {et~al.} 2011, ArXiv
  e-prints

\end{thebibliography}

\end{document}